\documentstyle[12pt]{article}
 \newcommand{\bea}{\begin{equation}}
 \newcommand{\eea}{\end{equation}}
 \newcommand{\ber}{\begin{eqnarray}}
 \newcommand{\eer}{\end{eqnarray}}
\begin{document}
\title{FREQUENCY DEPENDENT VISCOSITY NEAR THE CRITICAL POINT: THE SCALE TO TWO LOOP ORDER}
\author{Palash Das and Jayanta.K.Bhattacharjee\\
Department of Theoretical Physics \\
Indian Association for the Cultivation of Science \\
Jadavpur, Calcutta 700 032, India}
\date{}
\maketitle
\begin{abstract}
The recent accurate measurements of Berg, Moldover and Zimmerli of the viscoelastic effect near the critical point of xenon has shown that the scale factor involved in the frequency scaling is about twice the scale factor obtained theoretically. We show that this discrepancy is a consequence of using first order perturbation theory. Including two loop contribution goes a long way towards removing the discrepancy.
\vspace{1cm}

PACS number(s):64.60Ht 
\end{abstract}
\newpage
The shear viscosity $\eta $ of a liquid-gas system near the critical point or a binary liquid mixture near the critical consolute point has a weak divergence characterized by a small exponent $x_\eta$. If $\epsilon$ is the deviation of the temperature T from the critical temperature $T_c$,i.e $\epsilon$=(T-$T_c$)/T, then the correlation length $\xi$ of the fluctuation diverges as $\epsilon\rightarrow 0$, according to $\xi$ $\sim$ $\epsilon^{-\nu}$
The divergence of the shear viscosity is expressed as
\bea
\eta \propto \xi^{x_{\eta}} 
\eea
 If the measurements are conducted at a finite frequency , a new length scale enters the problem. This is due to critical slowing down , which implies that the time $\tau$ of the fluctuations of size $k^{-1}$ (k is the wave number of the fluctuation) diverges as $\tau \propto k^{-z}$ for small k. This produces a length scale $\tau^{1/z}$, which is $\omega^{-1/z}$ where $\omega$ is the frequency at which process is being probed. If the external frequency goes to zero, this length scale $l_\omega \propto \omega^{-1/z}$ goes to infinity and $\xi$ is the only controlling length in the problem. For a finite value of $l_\omega$ ,it is possible to go also enough to the critical point ,such that $\xi \sim l_{\omega}$ and then as $\xi$ exceeds $l_{\omega}$, the viscosity can change no longer and is determined by $l_{\omega}$. Thus for $\xi \:\rightarrow\ \infty$,
\bea
\eta \propto l_\omega^{x_{\eta}} \propto \omega^{-{x_\eta}/z}
\eea
In D-dimension ,the exponent z=D+$x_\eta$ and a finity values of $\xi$ and $\omega$ ,the viscosity is described by the scaling law (7,8),
\bea
\eta(\xi,\omega)=\xi^{x_{\eta}}[S(\omega/\Gamma_0\kappa^z)]^{-x_{\eta}/(D+x_{\eta})}
\eea
where $\Gamma_0\kappa^z$ is the characteristic frequency associated with the decay of fluctuations and S is a scaling function characterized  by S(0)=constant and S(y) $\propto$ y for y$>>$1. The above constraints on S(y) helps us recover Equations (1) and (2). In view of the smallness of the exponent $x_{\eta}$ ($\simeq$ 0.067), we can expand $\xi^{x_{\eta}}$ as 
1+$x_{\eta}ln{\xi}$ and writing 
$[S(y)]^{-{x_\eta}/D+{x_\eta}}$=1-$[x_{\eta}/(D+x_{\eta})]lnS(y)$+......,we arrive at 
\bea
\Delta \eta=\eta(\xi,\omega)-\eta(\xi,0)=-\eta(\xi,0)[x_{\eta}/(D+x_{\eta})]ln(\omega/\Gamma_0\kappa^z)
\eea
The simplest scaling function that one can think of is
\bea
ln S(\Omega)=ln(1+a\Omega)
\eea
where $\Omega = \omega/(2\Gamma_0\kappa^z) = \frac{\omega}{\frac{2kT}{6\pi\eta_0}\kappa^(3+x_{\eta})}$
which is the scaled frequency with the frequency scale set by the Kawasaki form and "a" is a number of O(1) which describes where the crossover from the "hydrodynamic" (zero frequency) to the "non-hydrodynamic" (frequency limited) behaviour takes place.
\vspace{.1cm}
\hspace{1cm}

In the one-loop self consistent calculation, the frequency dependent shear viscosity in D=3 is given by (see Fig.1)
\ber
\eta_1({\kappa,\omega})&=&\frac{1}{4}\int \frac{d^3p}{(2\pi)^3}\frac{4 p^4 sin^2{\theta}cos^2{\theta}}{(p^2+\kappa^2)^2[-i\omega+2\Gamma_0p^2\sqrt{p^2+\kappa^2}]}\nonumber \\
&=&\frac{1}{30\pi^2\Gamma_0}\int \frac{ p^6 dp}{(p^2+1)^2[-i\Omega+p^2\sqrt{p^2+1}]}
\eer
leading to
\bea
\Delta\eta({\kappa,\omega})=\frac{1}{30\pi^2\Gamma_0}\int \frac{p^6dp}{{(p^2+1)}^2}[\frac{1}{[-i\Omega+p^2\sqrt{p^2+1}]}-\frac{1}{[p^2\sqrt{p^2+1}]}]
\eea
Comparing with Equations (4)and (5) ,the function lnS($\Omega$)to one loop is ln$S_1$($\Omega$),given by
\bea
lnS_1(\Omega)=-\int \frac{p^6dp}{{(1+p^2)}^2}[\frac{1}{[-i\Omega+p^2\sqrt{p^2+1}]}-\frac{1}{[p^2\sqrt{p^2+1}]}]
\eea
which has the high frequency form, ln$S_1$($\Omega$)$\simeq$$\frac{1}{3}$ln$\Omega$/$e^{4-3ln2}$, corresponding to "a"=0.147. Yet another way of estimating "a" is to study the low frequency limit of Eq.(8), from where we find 
\bea
lnS_1(\Omega)=-i\Omega\frac{\pi}{16}+O(\Omega^2)
\eea
which corresponds to "a"=$\frac{3\pi}{16}$=0.589, when we compare with the low frequency Taylor expansion of Eq.(9). Thus, we see that for the crossover frequency scale "a", there are two possible estimates (8). One comes from the high frequency end, which we call $a_{h_i}$ and another from the low frequency end, which we call $a_{l_0}$. At one loop level $a_{h_i}$=0.147 and $a_{l_0}$=0.589. The two estimates are quite different.The full scaling function, obtained by evaluating the complete integral in Eq.(8), gives the gradual change in scale from $a_{l_0}$ to $a_{h_i}$. This was done in Ref.(8).
\vspace{.1cm}
\hspace{1cm}

In recent accurate viscoelastic effect measurements of Berg et al.(9,10), it was reported that the true frequency scale is about twice as big as the theoretical one. Since the one loop scaling function has a changing frequency scale, it is worthwhile examining this observed discrepancy more closely. Accordingly,we analysed the data of Berg et al.slightly differently. For the different t=$\frac{(T-T_c)}{T_c}$ values frequencies studied by Berg et al., we evaluate the dimensionless frequency $\Omega$=2$\pi$$\frac{3\pi\eta_0}{kT\kappa^(3+x_\eta)}$=$\pi$f$\tau_0$, where $\tau_0$=$\frac{kT\kappa^(3+x_\eta)}{6\pi\eta_0}$ is the decay rate for concentration fluctuations. We notice that more than 75$\%$ of the data of Berg et al. are in the range $\Omega$$<$1. In the range where $\Omega$ is small, The ratio R=$\frac{Im\eta(\kappa,\omega)}{Re\eta(\kappa,\omega)}$ is linear in $\Omega$. This ratio is the most direct probe of the viscoelastic effect and one needs to concentrate on it. It is clear that for $\Omega$$<<$1
\vspace{.05cm}
\hspace{3cm}

R=$\frac{x_\eta}{3+x_\eta}$$a_{l_0}$$\Omega$.
\vspace{.1cm}

The result of plotting R vs. $\Omega$ is shown in Fig.1. The one loop theory is shown by the dashed line.The slope of the data almost double. This is the discrepancy reported by Berg et al. In view of this differency, we have undertaken a two loop calculation of the frequency scale. In the low frequency end, we find a significant correction to the scale. This results in the solid line shown in Fig.1. In the high frequency end, the correction to the scale is similar. The data in the high frequency end is sparse. As far as we can tell, the scale that can be extracted from the data ($a_{h_i}$)is significantly smaller the low frequency result ($a_{l_0}$). This is consistent with the calculation. The remaining difference between the experimental slope at low frequencies and the calculation can be attributed to the loops left out .The fact that including two loop correction is a significant effect, is a clear out pointer to the importance of higher order terms in perturbation theory at this end. We now outline the calculation involved. It should be noted that in the self consistent scheme that we employ, the two loop graphs corresponding to self energy insertions in the propagating lines have already been taken into account at the dressed one loop level. Consequently,treating the vertex correction alone enables us to provide a complete two loop order calculation.
\vspace{.1cm}
\hspace{2cm}

The two loop contribution to the shear viscosity $\eta$ (see Fig.2)
is given by the vertex correction diagram and can be written as
\ber
\eta_2k^2&=&\frac{1}{D-1}\int \frac{d^Dp}{(2\pi)^D}\int
\frac{d^Dp}{(2\pi)^D}  \frac{[p^2-(\vec{k}-\vec{p})^2][q^2-(\vec{k}-\vec{q})^2]}{\eta_0(p^2+\kappa^2)(q^2+\kappa^2)\vert\vec{p}+\vec{q}-\vec{k}\vert^2}\nonumber \\
& &\frac{ (p_\alpha\tau_{\alpha\beta}(\vec{k})q_\beta)(p_\alpha\tau_{\alpha\beta}(\vec{k}-\vec{p}-\vec{q})q_{\beta})}
{[-i\omega+\Gamma(\vec{p},\kappa)+\Gamma(\vec{k}-\vec{p},\kappa)][-i\omega+\Gamma(\vec{q},\kappa)+\Gamma(\vec{k}-\vec{q},\kappa)]}\nonumber \\
& &{}
\eer
In the above $\tau_{\alpha\beta}(\vec{k})$ is the projection operator $\delta_{\alpha\beta}$-$\frac{k_\alpha k_\beta}{k^2}$and $\Gamma(k,\kappa)$  is the fully dressed order parameter relaxation rate.  From the right hand side, we need to extract the O($k^22$) term. We also need to average over all possible directions of $\vec{k}$.\\Accordingly,\\$p^2$-$(\vec{k}-\vec{p})^2$ $\simeq$ -2$\vec{k}.\vec{p}$\\$q^2$-$(\vec{k}-\vec{q})^2$ $\simeq$ -2$\vec{k}.\vec{q}$\\and\\$<(\vec{k}.\vec{p})(\vec{k}.\vec{q})p_\alpha\tau_{\alpha\beta}(\vec{k})q_\beta>$=$\frac{k^2p^2q^2[Dcos^2\theta-1]}{D(D+2)}$\\Everywhere else in the right hand side, we may set k=0 in the right hand side of Eq.(10). Thus the directional average of $p_\alpha\tau_{\alpha\beta}(\vec{k}-\vec{p}-\vec{q})q_\beta$ becomes\\$<p_\alpha\tau_{\alpha\beta}(\vec{p}+\vec{q})q_\beta>$ and can be written as\bea<p_\alpha\tau_{\alpha\beta}(\vec{p}+\vec{q})q_\alpha>=-\frac{p^2q^2sin^2\theta}{(\vec{p}+\vec{q})^2}
\eea
We can now write Eq.(10) as 
\ber
\eta_2&=&\frac{4}{(D-1)D(D+2)\eta_0}\int \frac{d^Dp}{(2\pi)^D}\int \frac{d^Dq}{(2\pi)^D}\frac{p^4q^4sin^2\theta[1-Dcos^2\theta]}{(p^2+\kappa^2)(q^2+\kappa^2)(\vec{p}+\vec{q})^4}\nonumber\\
& &\frac{1}{[-i\omega+2\Gamma(\vec{p},\kappa)][-i\omega+2\Gamma(\vec{q},\kappa)]}\nonumber\\
& &{}
\eer
Specializing to D=3,we can replace the relaxation rate $\Gamma(k,\kappa)$ by an accurate approximate to the full Kawasaki function as $\Gamma(k,\kappa)=\Gamma_0k^2(k^2+\kappa^2)^2$,and we have
\ber
\eta_2&=&\frac{1}{30\eta_0\Gamma_0\Gamma_0}\int \frac{d^3p}{(2\pi)^3}\int \frac{d^3q}{(2\pi)^3}\frac{p^4q^4sin^2\theta[1-3cos^2\theta]}{(p^2+1)(q^2+1)(\vec{p}+\vec{q})^4}\nonumber\\
& &\frac{1}{[-i\Omega+p^2\sqrt(1+p^2)][-i\Omega+q^2\sqrt(1+q^2)]}\nonumber\\
& &{}
\eer
The two loop contribution $lnS_2$ to lnS is accordingly
\ber
lnS_2&=&-\frac{4\pi^2}{\eta_0\Gamma_0}\int \frac{d^3p}{(2\pi)^3}\int \frac{d^3q}{(2\pi)^3}\frac{p^4q^4sin^2\theta[1-3cos^2\theta]}{(1+p^2)(1+q^2)(\vec{p}+\vec{q})^4}\nonumber\\
& &[\frac{1}{[-i\Omega+p^2\sqrt(1+p^2)][-i\Omega+q^2\sqrt(1+q^2)]}-\frac{1}{p^2\sqrt(1+p^2)q^2\sqrt(1+q^2)}]\nonumber\\
& &{}
\eer
Our first task is to ensure that the zero frequency integrals reproduce the correct two loop viscosity exponent $x_\eta$. Accordingly, from Equations(6)and(13), we find (the product $\eta_0\Gamma_0$ is fixed by the diffusion coefficient diagrammatics to be $\frac{1}{16}$)
\bea
\eta=\eta_1(\kappa,0)+\eta_2(\kappa,0) \simeq \eta_0\frac{8}{15\pi^2}(1+\frac{8}{3\pi^2})ln\frac{\Lambda}{\kappa}+.... 
\eea
The coefficient of $\eta_0ln\frac{\Lambda}{\kappa}$ in the above equation is within 2 \% (the correction coming from two loop self energy insertion graphs and the dissipative four point coupling)is the exponent $x_\eta$. We can now carry out a Taylor expansion of the right hand side of Eq.(14). This yields the two loop contribution $a_2$ to the scale factor $a_l{_0}$ as
\bea
a_2=\frac{16}{15}\int \frac{d^3p}{(2\pi)^3(1+p^2)^2}\int \frac{d^3q sin^2\theta(1-3cos^2\theta)q^2}{(2\pi)^3(1+q^2)^{\frac{3}{2}}(\vec{p}+\vec{q})^4}
\eea
which yields
\bea
a_{l_0}=3[\frac{\pi}{16}+\frac{8}{3\pi^2} \times\frac{2\pi}{3}\times 0.115]=0.78
\eea
Using the above scale, the solid line has been drawn in Fig.1. While this is a significant improvement over the one loop answer(dashed line), we still have an infinite number of loops left out and the combined effect should be to remove the remaining discrepancy.\\Turning to the high frequency side, evaluation of the appropriate integrals (Eq.13) show that the behaviour of zero frequency viscosity to two loop order is
\bea
\eta=\eta_0\frac{8}{15\pi^2}(1+\frac{8}{3\pi^2})[ln\frac{\Lambda}{\kappa}-\frac{(\frac{4}{3}-ln2)+\frac{8}{3\pi^2}\times0.85}{(1+\frac{8}{3\pi^2)}}]
\eea
while the high frequency limit is 
\bea
\eta=\eta_0\frac{8}{15\pi^2}(1+\frac{8}{3\pi^2)}[\frac{1}{3}ln\frac{\Lambda^3}{-i\Omega}-\frac{\frac{8}{3\pi^2}\times 0.90}{(1+\frac{8}{3\pi^2})}]
\eea
From Eqs.(18)and(19), the two loop $a_{h_i}$is
\bea
a_{h_i}=0.23
\eea
which is a 50\% increase over the one loop result.
\vspace{.1cm}
\hspace{2cm}

Thus,we see that the scale factor associated with the frequency dependent viscosity near the critical point undergoes a significant enhancement both in the high and low frequency range, when the perturbation theory is carried out to two loop order. This helps us the reported discrepancy between the measurements of Berg et al.and the one loop calculation.

\end{document}